\documentclass[a4paper,12pt]{article}
\usepackage[pctex32]{graphics}

\begin{document}
\topmargin 0pt \oddsidemargin 0mm

\renewcommand{\thefootnote}{\fnsymbol{footnote}}
\begin{titlepage}
\begin{flushright}
INJE-TP-05-07\\
gr-qc/0509040
\end{flushright}

\vspace{5mm}
\begin{center}
{\Large \bf Equation of state for an interacting holographic dark
energy model} \vspace{12mm}

{\large  Hungsoo Kim$^1$, Hyung Won Lee$^{1}$  and Yun Soo
Myung$^{1,2}$\footnote{e-mail
 address: ysmyung@physics.inje.ac.kr}}
 \\
\vspace{10mm} {\em  $^{1}$Relativity Research Center, Inje
University,
Gimhae 621-749, Korea \\
$^{2}$Institute of Theoretical Science, University of Oregon,
Eugene, OR 97403-5203, USA}

\end{center}

\vspace{5mm} \centerline{{\bf{Abstract}}}
 \vspace{5mm}
We investigate  a model of the interacting holographic dark energy
with cold dark matter (CDM).  If the holographic energy density
decays into CDM, we find two types of the effective equation of
state. In this case we have to use the effective equations of
state ($\omega^{\rm eff}_{\rm \Lambda}$) instead of the equation
of state ($\omega_{\rm \Lambda})$. For a fixed ratio of two energy
densities, their effective equations of state  are given by the
same negative constant. Actually, the cosmic anti-friction arisen
from the vacuum decay process may induce the acceleration with
$\omega^{\rm eff}_{\rm \Lambda}<-1/3$. For a variable ratio, their
effective equations of state are slightly different, but they
approach the same negative constant in the far future.
Consequently, we show that such an interacting holographic energy
model cannot accommodate a transition from the dark energy with
$\omega^{\rm eff}_{\rm \Lambda}\ge-1$ to the phantom regime with
$\omega^{\rm eff}_{\rm \Lambda}<-1$.
\end{titlepage}
\newpage
\renewcommand{\thefootnote}{\arabic{footnote}}
\setcounter{footnote}{0} \setcounter{page}{2}

\section{Introduction}
Supernova (SN Ia) observations suggest that our universe is
accelerating and the dark energy contributes $\Omega_{\rm
DE}\simeq 0.75$ to the critical density of the present
universe~\cite{SN}. Also  cosmic microwave background
observations~\cite{Wmap} imply that the standard cosmology is
given by the inflation and FRW universe~\cite{Inf}. Although there
exist a number of dark energy candidates, the two candidates are
the cosmological constant  and the quintessence scenario. The
equation of state (EOS) for the latter is determined dynamically
by the scalar or tachyon. In the study of dark energy~\cite{UIS},
the first issue is whether or not the dark energy is a
cosmological constant with $\omega_{\Lambda}=-1$. If the dark
energy is shown not to be a cosmological constant, the next  is
whether or not the phantom-like state of $\omega_{\Lambda}<-1$ is
allowed. Most theoretical  models that  can  explain
$\omega_{\Lambda}<-1$ confront with serious problems. The last one
is whether or not $\omega_{\Lambda}$ is changing as the universe
evolves.

On the other hand, there exists another model of the dynamical
cosmological constant derived  by the holographic principle. The
authors in~\cite{CKN} showed that in quantum field theory, the UV
cutoff $\Lambda$ could be  related to the IR cutoff $L_{\rm
\Lambda}$ due to the limit set by introducing  a black hole. In
other words, if $\rho_{\rm \Lambda}=\Lambda^4$ is the vacuum
energy density caused by the UV cutoff, the total energy of system
with the size $L_{\rm \Lambda}$ should not exceed the mass of the
black hole with the same size $L_{\rm \Lambda}$ : $L_{\rm
\Lambda}^3 \rho_{\rm \Lambda}\le 2L_{\rm \Lambda}/G$. The
newtonian constant $G$ is given by the Planck mass ($G=1/M_p^2$).
If the largest cutoff $L_{\rm \Lambda}$ is chosen to be the one
saturating this inequality,  the holographic energy density is
then given by $\rho_{\rm \Lambda}= 3c^2M_p^2/8\pi L_{\rm
\Lambda}^2$ with an undetermined constant $c$. Here we regard
$\rho_{\rm \Lambda}$ as the dynamical cosmological constant.
Taking $L_{\rm \Lambda}$ as the size of the present universe
(Hubble horizon: $R_{\rm HH}$), the resulting energy  is close to
the present dark energy~\cite{HMT}. Even though it may explain the
data, this
 approach is not complete. This is because it fails to recover the EOS for
a dark energy-dominated universe~\cite{HSU}.

 Usually, it is
not an easy matter to determine the equation of state for  a
system with UV/IR cutoff. In order to find the EOS, we propose the
two approaches. Firstly,  the future event horizon of $R_{\rm FH}$
is used for the IR cutoff $L_{\rm \Lambda}$ instead of $R_{\rm
HH}$~\cite{LI}. In this case, one  finds that $\rho_{\rm \Lambda}
\sim a^{-2(1-1/c)}$. It may describe the dark energy with
$\omega_{\rm \Lambda}=-1/3-2/3c~(c\ge 1)$. For example, one
obtains $\omega_{\rm \Lambda}=-1$ for $c=1$. The related issues
appeared in Ref.~\cite{FEH,Myung2}. Secondly, one may introduce an
interaction between the holographic energy density with $R_{\rm
HH}$ and CDM. Here the EOS for the holographic energy density is
less important because the interaction changes it~\cite{Zim3}.
Recently, the authors in \cite{WGA} introduced an interacting
holographic dark energy model. They derived the phantom-like EOS
of $\omega_{\rm \Lambda}<-1$ for a model that an interaction
exists between holographic energy  with $L_{\rm \Lambda}=R_{\rm
FH}$ and CDM. They insisted that this model can describe even the
phantom regime with $\omega_{\rm \Lambda}<-1$. This implies that
the interacting holographic model can accommodate a transition of
the dark energy from a normal state
 to a phantom regime.
Although the decay process leads to the case  that the effective
EOS of CDM becomes negative, but this process does not change the
nature of holographic energy into the phantom-like matter
significantly. Hence it  is hard to accept their argument because
they consider the process of decaying from the holographic energy
density  into CDM.

In this work  we examine this issue carefully.  We will show that
 the interacting holographic dark energy model cannot describe a
phantom regime of $\omega^{\rm eff}_{\rm \Lambda}<-1$ when using
the effective EOS.  A key of this  system is an interaction
between holographic energy and CDM. Their contents are changing
due to energy transfer from holographic energy to CDM until the
two components are comparable. If there exists a source/sink in
the right hand side of the continuity equation, we must be careful
to define the  EOS. In this case the effective EOS is the only
candidate to represent the state of the mixture of two components
arisen from decaying of the holographic energy into CDM. This is
quite
 different from the non-interacting case. Hence we remark
an important usage which is useful for our study
\begin{eqnarray}
\label{1eq1}&& {\rm effective~ EOS} \Longrightarrow {\rm an~ interacting ~two~fluid~ model}, \nonumber \\
\label{1eq2}&& {\rm  EOS} \Longrightarrow {\rm a ~noninteracting~
two~fluid~ model} \nonumber.
\end{eqnarray}

\section{Interacting model}
 Let us imagine a universe made of CDM with
$\omega_{\rm m}=0$, but obeying the holographic principle. In
addition, we propose that the holographic energy density  exists
with $\omega_{\rm \Lambda}\ge-1$. If one introduces a form of the
interaction $Q=\Gamma \rho_{\rm \Lambda}$, their continuity
equations take the forms
\begin{eqnarray}
\label{2eq1}&& \dot{\rho}_{\rm \Lambda}+3H(1+\omega_{\rm \Lambda})\rho_{\rm \Lambda} =-Q, \\
\label{2eq2}&& \dot{\rho}_{\rm m}+3H\rho_{\rm m}=Q.
\end{eqnarray}
 This implies that the mutual interaction could
provide a mechanism to the particle production. Actually this is a
decaying of the holographic energy component into CDM with the
decay rate $\Gamma$.  Taking a ratio of two energy densities as
$r=\rho_{\rm m}/\rho_{\rm \Lambda}$, the above equations lead to
\begin{equation}
\label{2eq3} \dot{r}=3Hr\Big[\omega_{\rm \Lambda}+
\frac{1+r}{r}\frac{\Gamma}{3H}\Big] \end{equation} which means
that the evolution of the ratio depends on the explicit form of
interaction. In this work we choose the same notation as in
Ref.~\cite{WGA}, $\Gamma=3b^2(1+r)H$ with  the coupling constant
$b^2$. Even if one starts with $\omega_{\rm m}=0$ and $\omega_{\rm
\Lambda}=-1$, this process is necessarily accompanied by the
different equations
 of state  $\omega^{\rm eff}_{\rm m}$ and $\omega^{\rm eff}_{\rm \Lambda}$. The decaying process impacts
their equations of state  and particularly, it provides the
negatively effective EOS of CDM.  Actually, an  accelerating phase
could arise from a largely effective non-equilibrium pressure
 $\Pi_{\rm m}$  defined as $\Pi_{\rm m}\equiv -\Gamma\rho_{\rm \Lambda}/3H(\Pi_{\rm
 \Lambda}=\Gamma\rho_{\rm \Lambda}/3H)$. Then  the two  equations (\ref{2eq1})
and (\ref{2eq2}) are translated into those of the two
dissipatively imperfect fluids
\begin{eqnarray}
\label{2eq4}&& \dot{\rho}_{\rm \Lambda}+ 3H\Big[1+\omega_{\rm
\Lambda}+\frac{\Gamma}{3H} \Big]\rho_{\rm \Lambda}=\dot{\rho}_{\rm
\Lambda}+ 3H\Big[(1+\omega_{\rm \Lambda})\rho_{\rm
\Lambda}+\Pi_{\rm
 \Lambda}\Big]=0, \\
\label{2eq5}&& \dot{\rho}_{\rm
m}+3H\Big[1-\frac{1}{r}\frac{\Gamma}{3H}\Big]\rho_{\rm
m}=\dot{\rho}_{\rm m}+3H(\rho_{\rm m}+\Pi_{\rm m})=0.
\end{eqnarray}
$\Pi_{\rm \Lambda}>0$ shows a decaying  of holographic energy
density via the cosmic frictional force, while $\Pi_{\rm m}<0$
induces a production of the CDM via the cosmic anti-frictional
force simultaneously~\cite{Zim1,myung}. This is a sort  of the
vacuum decay process to generate a particle production within the
two fluid model~\cite{Zim2}. As a result, a mixture of two
components will be  created. From Eqs.(\ref{2eq4}) and
(\ref{2eq5}), turning on the interaction term, we define their
effective equations of state as
\begin{equation}
\label{2eq6} \omega^{\rm eff}_{\rm \Lambda}=\omega_{\rm
\Lambda}+\frac{\Gamma}{3H},~~ \omega^{\rm eff}_{\rm
m}=-\frac{1}{r}\frac{\Gamma}{3H}. \end{equation}

On the other hand, the first Friedmann equation is given by
\begin{equation}
\label{2eq7} H^2=\frac{8\pi}{3M^2_p}\Big[
 \rho_{\rm \Lambda}+\rho_{\rm m}\Big].
\end{equation}
 Differentiating
Eq.(\ref{2eq7}) with respect to the cosmic time $t$ and then using
Eqs.(\ref{2eq1}) and (\ref{2eq2}), one finds the second Friedmann
equation as\footnote{It seems that the deceleration parameter of
$q=-1-\dot{H}/H^2$ is independent of the interaction factor
$\Gamma(\sim b^2)$. However, using Eq.(\ref{3eq4}), one finds that
$q=1/2-3b^2/2-\Omega_{\rm \Lambda}/2-\Omega^{3/2}_{\rm \Lambda}/c$
\cite{WGA}. Even for $r=$const, using Eq.(\ref{2eq3}) leads to
$\omega_{\rm \Lambda}=-b^2(1+r)^2/r$. This means that the
acceleration will be determined by $\dot{H}$ through $\omega_{\rm
\Lambda}$.}
\begin{equation}
\label{2eq8} \dot{H}=- \frac{3}{2}H^2\Big[1+\frac{
 \omega_{\rm \Lambda}}{1+r}\Big].
\end{equation}
 Let us
introduce
\begin{equation}
\label{2eq9} ~\Omega_{\rm m}=\frac{8\pi \rho_{\rm
m}}{3M_p^2H^2},~\Omega_{\rm \Lambda}=\frac{8 \pi \rho_{\rm
\Lambda}}{3M^2_pH^2} \end{equation} which allows to rewrite the
first Friedmann equation as
\begin{equation} \label{2eq10} \Omega_{\rm m}+\Omega_{\rm
\Lambda}=1.\end{equation}
 Then we can express $r$ and
its derivative ($\dot{r}$) in terms of $\Omega_{\rm \Lambda}$ as
\begin{equation}
\label{2eq11} r=\frac{1-\Omega_{\rm \Lambda}}{\Omega_{\rm
\Lambda}},~~ \dot{r}=-\frac{\dot{\Omega}_{\rm
\Lambda}}{\Omega^2_{\rm \Lambda}}.
\end{equation}
 Here we get an important relation of $\Omega_{\rm \Lambda}=1/(1+r)$
  between $\Omega_{\rm \Lambda}$ and $r$.

\section{Holographic energy density with the future event horizon}
In the case of $\rho_{\rm \Lambda}$ with $L_{\rm \Lambda}=1/H$, we
always have a fixed ratio of two energy densities. This provides
the same negative EOS for both two components~\cite{Zim3,myung}.
In order to study a variable ratio of two energy densities, we
need to introduce the future event horizon~\cite{LI,FEH}
\begin{equation}
\label{3eq1}
 L_{\rm \Lambda}=R_{\rm FH} \equiv a \int_t^{\infty}
(dt/a)=a \int_a^{\infty}(da/Ha^2). \end{equation}
 In this case the
first Friedmann equation takes the form (\ref{2eq7}) with
$\rho_{\rm \Lambda}=\frac{3 c^2M^2_p}{8\pi R^2_{\rm FH}}$. From
this we derive a reduced equation
\begin{equation}
\label{3eq2} R_{\rm
FH}=\frac{c\sqrt{1+r}}{H}=\frac{c}{H\sqrt{\Omega_{\rm
\Lambda}}}.\end{equation}  Considering  the definition of
holographic energy density $\rho_{\rm \Lambda}$, one finds also
\begin{equation} \label{3eq3}
\dot{\rho}_{\rm \Lambda}=2H\rho_{\rm \Lambda}\Big[-1+
\frac{1}{R_{\rm FH}H}\Big]=-3H\rho_{\rm \Lambda}\Big[1-\frac{1}{3}
-\frac{2\sqrt{\Omega_{\rm \Lambda}}}{3c}\Big].
\end{equation}
It can be easily integrated to give $\rho_{\rm \Lambda} \sim
a^{-3(1+\omega^{\rm eff}_{\rm \Lambda})}$ with $\omega^{\rm
eff}_{\rm \Lambda}=-1/3- 2\sqrt{\Omega_{\rm \Lambda}}/3c$ only for
$r$=const~($\Omega_{\rm \Lambda}$=const). On the other hand,
differentiating Eq.(\ref{3eq2}) with respect to the  cosmic time
$t$ leads to two important relations. Using Eqs.(\ref{2eq3}) and
(\ref{2eq8}), one finds the holographic energy equation of state
\begin{equation} \label{3eq4}
\omega_{\rm \Lambda}=-\frac{1}{3}-\frac{2\sqrt{\Omega_{\rm
\Lambda}}}{3c}-\frac{b^2}{\Omega_{\rm \Lambda}}.
\end{equation}
The other is cast in a form of differential equation for
$\Omega_{\rm \Lambda}$
\begin{equation} \label{3eq5}
\frac{1}{\Omega^2_{\rm \Lambda}}\frac{d \Omega_{\rm
\Lambda}}{dx}=(1-\Omega_{\rm \Lambda})\Big[\frac{1}{\Omega_{\rm
\Lambda}}+\frac{2}{c\sqrt{\Omega_{\rm
\Lambda}}}-\frac{3b^2}{\Omega_{\rm \Lambda}(1-\Omega_{\rm
\Lambda})}\Big]
\end{equation}
with $x=\ln a$. Plugging the solution to Eq.(\ref{3eq5}) into
Eq.(\ref{3eq4}), one can determine the evolution of equation of
state. These equations were also derived in Ref.~\cite{WGA}.

As a simple example, we first consider a fixed ratio of two energy
densities.
 In this case of $r$=const, we obtain from Eq.(\ref{2eq3})
\begin{equation}
\label{3eq6} \omega_{\rm \Lambda}=-
\frac{1+r}{r}\frac{\Gamma}{3H}=-\frac{b^2}{\Omega_{\rm
\Lambda}(1-\Omega_{\rm \Lambda})} \end{equation} which means that
$\omega_{\rm \Lambda}=0$, if there is no interaction ($\Gamma=0$).
Substituting  this into Eq.(\ref{2eq6}), one obtains   the same
effective EOS for both components
\begin{equation}
\label{3eq7} \omega^{\rm eff}_{\rm
\Lambda}=-\frac{b^2}{1-\Omega_{\rm \Lambda}}=\omega^{\rm eff}_{\rm
m}. \end{equation} Furthermore, from Eq.(\ref{3eq5}) one finds a
relation which is valid for $\Omega_{\rm \Lambda}$=const
\begin{equation}
\label{3eq8} 1-\frac{\sqrt{\Omega_{\rm
\Lambda}}}{c}=\frac{3}{2}\Big(1-\frac{b^2}{1-\Omega_{\rm
\Lambda}}\Big). \end{equation} Using the above relation, one
arrives at
\begin{equation}
\label{3eq9} \omega^{\rm eff}_{\rm
\Lambda}=-\frac{1}{3}-\frac{2\sqrt{\Omega_{\rm
\Lambda}}}{3c}=\omega^{\rm eff}_{\rm m}. \end{equation}

We confirms from Eq. (\ref{3eq3}) that the effective equation
 of state (\ref{3eq9}) is correct. This is very similar to
the case that the Hubble horizon is chosen for  the IR cutoff.
Using another notation of $\omega^{\rm eff}_{\rm
\Lambda}=\omega_{\rm \Lambda}/(1+r)$, one finds the same
expression as in the case found for the Hubble
horizon~\cite{Zim3}.  At this stage we emphasize  that in the
presence of interaction,  the true equation of state for the
holographic energy density is given by not $\omega_{\rm \Lambda}$
but $\omega^{\rm eff}_{\rm \Lambda}$.

\begin{figure}[t!]
   \centering
   \includegraphics{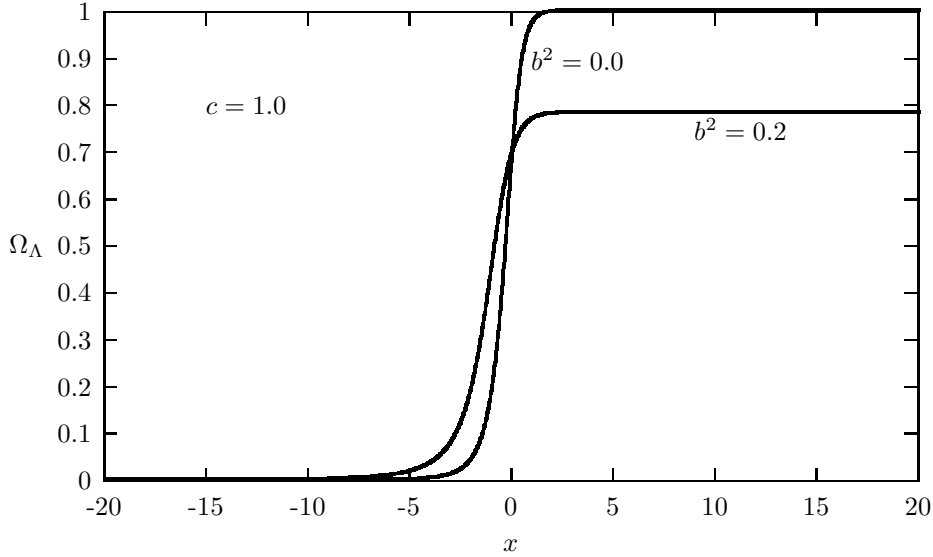}
   \caption{The evolution of density parameter $\Omega_{\rm \Lambda}(x)$
    as a monotonically increasing function of $x=\ln a$.
   Here we choose $c=1.0$ and $b^2=0.2$ for an interacting case,
   while $c=1.0, b^2=0$ for a noninteracting case. For the
   noninteracting case, it shows that $\Omega_{\rm \Lambda}(x) \to 1$
   as $x$ increases, but for the
   interacting case $\Omega_{\rm \Lambda}(x) \to 0.8$ as $x$
   increases.
   The latter is possible because two components
   become comparable after the interaction.}
   \label{fig1}
\end{figure}

 Now we are in a position to discuss  a variable ratio of two energy densities.
  From Eqs.(\ref{2eq6}) and (\ref{3eq4}), we have  the effective equation of state
\begin{equation}
\label{3eq10} \omega^{\rm eff}_{\rm \Lambda}(x)= \omega_{\rm
\Lambda}(x)+\frac{b^2}{\Omega_{\rm
\Lambda}(x)}=-\frac{1}{3}-\frac{2\sqrt{\Omega_{\rm
\Lambda}(x)}}{3c}.
\end{equation}
It seems that $\omega^{\rm eff}_{\rm \Lambda}(x)$ is independent
of the decay rate $\Gamma$. However,
 a solution $\Omega_{\rm \Lambda}(x)$ to the evolution equation (\ref{3eq5}) which includes the
 $b^2$-term
determines how the effective equation of state $\omega^{\rm
eff}_{\rm \Lambda}(x)$  is changing  under the evolution of the
universe. In this process the interaction impacts on both  the
holographic energy density and the CDM.  Accordingly, their
contents are changing due to energy transfer from the holographic
energy to the CDM until two components are comparable. As is shown
Fig. 1, $\Omega_{\rm \Lambda}(x)$ is  a monotonically increasing
function of $x=\ln a$. For the
   noninteracting case of $b^2=0$, we find that $\Omega_{\rm \Lambda}(x) \to 1$
   as $x$ increases, while  for the interaction case of $b^2=0.2$,  $\Omega_{\rm \Lambda}(x)
   \to 0.8$. The first case is obvious because the holographic energy   with
   the future event horizon dominates in the future. Further the latter shows that  two components
   become comparable, due to the interaction.

On the other hand, the effective equation of state for CDM is
given differently by
\begin{equation} \label{3eq11}
\omega^{\rm eff}_{\rm m}(x)=-\frac{b^2}{1-\Omega_{\rm
\Lambda}(x)}.
\end{equation}
 This arises because a relation of Eq.(\ref{3eq9}) is no longer valid
for the dynamic evolution of a variable ratio.

\begin{figure}[t!]
   \centering
   \includegraphics{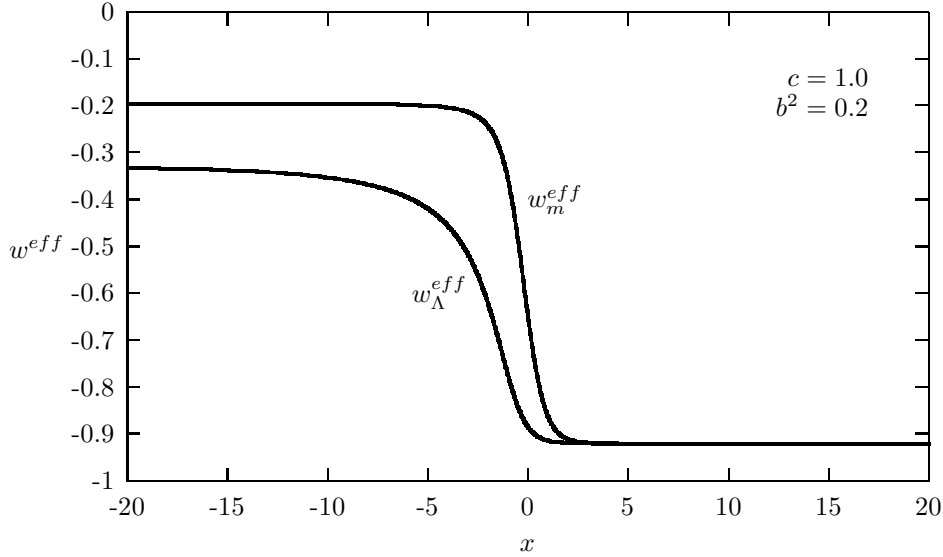}
   \caption{The effective equations of state for holographic energy and CDM versus $x=\ln a$.
   Here we choose $c=1$ and $b^2=0.2$ for simplicity. Although the two effective
   EOS show different behaviors during  evolution of the universe,
   these approach shortly the same value which is larger than $-1$ in the  future.
   This is possible because the two components
   become comparable after the interaction.}
   \label{fig2}
\end{figure}
We could conjecture the lower bound of $\omega^{\rm eff}_{\rm
\Lambda}(x)$ by requiring the holographic principle. According to
this principle, the total entropy $S=S_{\rm m}+S_{\rm \Lambda}$ of
the universe is bounded by the Bekenstein-Hawking entropy of
$S_{\rm BH}=\pi L^2_{\rm \Lambda}$. Here we choose $L_{\rm
\Lambda}=R_{\rm FH}=c/H\sqrt{\Omega_{\Lambda}}$. That is, one has
$S \le S_{BH}$. For simplicity, we assume that the entropy of the
universe is given roughly by the one saturating the bound ($S \sim
S_{BH}$). If one requires the second law of thermodynamics (the
entropy of the universe does not decrease, as the universe
evolves), one has a relation of  $\dot{S}_{BH} \ge 0$ which gives
$\dot{R}_{\rm FH}=c/\sqrt{\Omega_{\Lambda}}-1\ge 0$~\cite{LI}.
This implies that $c\ge \sqrt{\Omega_{\Lambda}}$. Applying this to
Eq. (\ref{3eq10}) leads to the lower bound: $\omega^{\rm eff}_{\rm
\Lambda}(x)\ge-1$. Accordingly it seems to be impossible to have
$\omega^{\rm eff}_{\rm \Lambda}(x)$ crossing $-1$. That is, the
phantom-like equation of state ($\omega^{\rm eff}_{\rm
\Lambda}(x)<-1$) is not allowed, even if one includes an
interaction between the holographic energy density and CDM.   This
feature can be confirmed from the  numerical computations using
Eqs.(\ref{3eq5}) and (\ref{3eq10}) (see Fig. 2). It  shows that
the effective EOS of $\omega_{\Lambda}^{eff}$ for the holographic
energy is always larger than $-1$ during the whole evolution of
the universe. As was shown at Fig. 5 in Ref.\cite{WGA},
$\omega_{\rm \Lambda}=\omega^{\rm eff}_{\rm
\Lambda}-b^2/\Omega_{\rm \Lambda}(x)$ is smaller than $-1$ in the
far future. In this case, however,  we have to use $\omega^{\rm
eff}_{\rm \Lambda}$ instead of $\omega_{\rm \Lambda}$ for a
description of the interacting case.

Finally, we wish to comment on the following case. One may require
that $\omega_{\rm \Lambda}$ itself be larger than $-1$, since the
holographic principle is compatible with  the dominant energy
condition of $\rho_{\rm \Lambda}\ge |p_{\rm \Lambda}|$. In this
case, we have $\omega_{\rm \Lambda}\ge-1$ and thus it may provide
the upper bound on the parameter $b^2$. This condition may work
for the noninteracting picture. However,  we have to use
$\omega^{\rm eff}_{\rm \Lambda}$ for the interacting picture. The
reason  is clear because the interaction makes a mixture of two
fluid which is different from CDM and holographic energy.  If one
requires this dominant energy condition on this mixture instead,
then one finds the known bound of $\omega^{\rm eff}_{\rm
\Lambda}\ge-1$, which is already obtained  by imposing the entropy
relation.

\section{Discussions}

We discuss a few of  pictures of the  vacuum decay in cosmology.
We usually introduce a source/sink to mediate an interaction
between holographic energy and CDM  in the continuity
equations~\cite{Zim1}. This picture is called the decaying vacuum
cosmology which may be related to the vacuum
fluctuations~\cite{PAD}. Here we wish to describe three different
pictures.

The first picture is that the  equation of state is fixed by
$p_{\rm \Lambda}=-\rho_{\rm \Lambda}$ for all time~\cite{WM}. As a
result of decaying the holographic energy  into the CDM, the
energy density of CDM takes a different form of $\rho_{\rm m} \sim
a^{-3+\epsilon}$ with a positive constant $\epsilon$. This means
that CDM will dilute more slowly compared to its standard form of
$\rho_{\rm m} \sim a^{-3}$.  However, this picture seems to  focus
on the CDM sector.

The second  is that the EOS for $\rho_{\rm \Lambda}$ is
indeterminate in the beginning but a ratio of two energy densities
is fixed. In this case  the holographic energy  itself is changing
as a result of decaying into the CDM.  Requiring the total
energy-momentum conservation, its change must be compensated by
the corresponding change in the CDM sector~\cite{Zim3}.
  The
two matters turn into  the imperfect fluids. The decaying process
continues until two components  are comparable. Here we note that
the effective EOS for the holographic energy and CDM  will be  the
same negative constant  by the interaction.   In this sense, the
works in~\cite{HOV,myung} are between the first picture and second
one, because they set $\omega_{\rm \Lambda}=-1$ initially and
determine $\omega^{\rm eff}_{\rm \Lambda}=-\epsilon/3=\omega^{\rm
eff}_{\rm m}$  with $L_{\rm \Lambda}=1/H$ or $R_{\rm FH}$ finally.

The third picture corresponds to the case that a ratio of two
energy densities is  changing as the universe evolves
~\cite{HSU,LI}. It works well for the presence of both the
holographic energy and CDM without interaction. In this case the
energy-momentum conservation is required for each matter
separately~\cite{LI}.
 Recently, it was proposed that this
picture is valid even for the case including an interaction
between the holographic energy with $R_{\rm FH}$ and CDM
\cite{WGA}. They used $\omega_{\rm \Lambda}$ to show that
$\rho_{\rm \Lambda}$  can describe the phantom regime. However, we
have to use $\omega^{\rm eff}_{\rm \Lambda}$  when considering the
interaction.  As are shown in Fig. 2, two equations of state take
different forms initially.  However, two effective EOS will take
the same negative value which is larger than $-1$ in the far
future.

Hence, the  vacuum decay picture is still alive even for a
dynamical evolution in the interacting holographic dark energy
model.  This implies that one  cannot  generate a phantom-like
mixture  of $\omega^{\rm eff}_{\rm \Lambda}<-1$ from an
interaction between the holographic energy and CDM. In other
words, decaying from the holographic energy  into  the CDM never
leads to the phantom regime.  Fig. 1 shows clearly that the
density parameter of  holographic energy is decreased from 1 to
0.8 when introducing an interaction of $b^2=0.2$. Furthermore,
from the graphs in Fig. 2, one recognizes the  changes from the
noninteracting case to the interacting one in the far future:
$\omega_{\rm \Lambda}=-1.0 \to \omega^{\rm eff}_{\rm
\Lambda}=-0.9$ and $\omega_{\rm m}=0 \to \omega^{\rm eff}_{\rm
m}=-0.9$. This means that although the CDM was changed
drastically, the holographic energy density preserves its nature.

Consequently, it is not true that after an inclusion of  the
interaction, the holographic energy density can describe the
phantom regime.

\section*{Acknowledgment}
Y. Myung thanks Steve Hsu,  Roman Buniy, and Brian Murray for
helpful discussions. H. Kim and H. Lee were  in part supported by
KOSEF, Astrophysical Research Center for the Structure and
Evolution of the Cosmos. Y. Myung  was in part supported by the
SRC Program of the KOSEF through the Center for Quantum Spacetime
(CQUeST) of Sogang University with grant number R11-2005-021.

\end{document}